# Bayesian Variable Selection with Related Predictors

Hugh Chipman

Graduate School of Business
University of Chicago
1101 E. 58th Street
Chicago, Illinois, 60637

hugh.chipman@gsb.uchicago.edu

August 1994, revised September 1995

## Abstract

In data sets with many predictors, algorithms for identifying a good subset of predictors are often used. Most such algorithms do not account for any relationships between predictors. For example, stepwise regression might select a model containing an interaction AB but neither main effect A or B. This paper develops mathematical representations of this and other relations between predictors, which may then be incorporated in a model selection procedure. A Bayesian approach that goes beyond the standard independence prior for variable selection is adopted, and preference for certain models is interpreted as prior information. Priors relevant to arbitrary interactions and polynomials, dummy variables for categorical factors, competing predictors, and restrictions on the size of the models are developed. Since the relations developed are for priors, they may be incorporated in any Bayesian variable selection algorithm for any type of linear model. The application of the methods here is illustrated via the Stochastic Search Variable Selection algorithm of George and McCulloch (1993), which is modified to utilize the new priors. The performance of the approach is illustrated with two constructed examples and a computer performance dataset.

# 1 Introduction

In regression models with numerous predictors, the issue of variable selection arises naturally. Given a set of $p$ predictor variables, the basic goal is to identify a "good" subset, using some appropriate criterion. Quite often, automatic techniques are used when $p$ is large, since there are a total of $2^p$ possible models from which to choose. When various types of predictors are included, the challenges become greater. For example, if interactions are considered, there are suddenly many more terms in the model, in addition to preconceptions about what types of models are good and what relations exist between predictors.

Typically, all $2^p$ models are either considered, or could be considered by these automated techniques. Some of these models are not likely to be used, even if they provide the best fit for the data. For example, some statisticians adopt the convention that a model containing an interaction should also contain the corresponding main effects. A common justification of this convention is that the models are easier to interpret. It is common practice to "clean up" the best model by adding in main effects corresponding to interactions in the model, or remove terms which are too complicated or difficult to explain. Not only is this technique time consuming, but it is done after the fact, rather than as an integrated part of the model selection procedure.

In doing so, statisticians are informally enforcing beliefs and preconceptions about which subsets of variables are likely to provide a good model. These beliefs implicitly specify relationships between predictors which are not usually recognized by automatic selection procedures. Possible reasons why these constraints are not built into subset selection algorithms include the additional bookkeeping necessary, and the inflexibility of the resulting procedures. The classification of models into possible and impossible groups is too coarse, and one may be reluctant to consign models to the latter class unless it is quite clear that they are not conceivable.

This paper proposes a richer, more mathematical way of expressing beliefs about the relationships between predictor variables. Instead of classifying a model as either "allowed" or "forbidden", a Bayesian approach is used to assign degrees of belief to each model in a structured way that accounts for relationships between predictors.

These generalizations deal with several different types of common dependencies between model terms. The first relates to two-way (and more general) interactions, and expresses the probability that an interaction will be active conditionally on whether its parents are active. A second type of relation groups like terms together and considers only models with all or none of these terms active. Relations between terms which are mutually exclusive are developed, and can be applied to selection of transformations for predictors, and to testing mutually exclusive hypotheses. Finally, global constraints on the model are introduced as a method of re-weighting models based on "larger scale" properties, such as the total number of active terms.

The priors developed can be used in any variable selection situation involving some sort of linear model. In what follows, the Stochastic Search Variable Selection (SSVS) algorithm of George and McCulloch (1993), henceforth referred to as GM, will be modified for these priors. This specific algorithm is chosen because of the large number of predictors considered and the all subsets nature of the search. Other methods, including the Markov Chain Monte Carlo Model Composition (MC$^3$) approach of Raftery, Madigan, and Hoeting (1993) could be used.

The next section presents an overview of SSVS and related matters. Priors that assign degrees of belief to different models are developed in Section 3, including interactions and higher order polynomials, grouped predictors, competing predictors, and restrictions on the number of predictors in the model. The behavior of these methods is illustrated via several constructed examples in Section 4. Section 5 presents the analysis of a computer performance data set, and introduces a graphic for examining the posterior.

## 2  Stochastic Search Variable Selection

This section reviews the SSVS algorithm of GM, which is based on the Gibbs sampler (see Smith and Roberts (1993) and references therein for an overview). It will be assumed that the criterion of interest is the posterior probability of a model conditional on the data. The approach in GM can be outlined as follows for the simplest case of linear regression with normal errors,

$$Y = X'\boldsymbol{\beta} + \epsilon. \qquad (1)$$

The central concept is to introduce an unobserved vector $\boldsymbol{\delta}$ of zeros and ones of the same length as $\boldsymbol{\beta}$. The components of this vector represent the importance of the corresponding regressor variables. That is, if $\delta_i = 0$, then the magnitude of $\beta_i$ is small, and the corresponding predictor is "inactive". If $\delta_i = 1$, then the magnitude of $\beta_i$ is large, and the predictor is "active". Mathematically, this is accomplished by defining a mixture prior for the coefficients $\boldsymbol{\beta}$:

$$f(\beta_i|\delta_i) = \begin{cases} N(0, \tau_i^2) & \text{if } \delta_i = 0 \\ N(0, (c_i\tau_i)^2) & \text{if } \delta_i = 1 \end{cases} \qquad (2)$$

When $\delta_i = 0$, $\beta_i$ is tightly centered around 0, and will not have a large effect. The much larger variance ($c_i \gg 1$) allows the possibility of a variable having a large influence. The parameters $c_i$ and $\tau_i$ must be chosen to represent a "small" effect, and how many times larger a "large" effect would be. The choice of appropriate values is important; recommendations given in GM are followed here.

This specific parameterization is chosen so that a Gibbs sampling approach may be used to obtain the posterior for $\boldsymbol{\delta}$. The basic idea of the Gibbs sampler is to construct a Markov chain whose state space is the parameter space, and whose equilibrium distribution is the joint distribution of the parameters (in a Bayesian context, the posterior). The Gibbs sampler constructs such a chain by successively sampling from conditional distributions. In GM, the conditional distributions are given by $f(\boldsymbol{\beta}|\boldsymbol{\delta}, \boldsymbol{\sigma}, Y), f(\boldsymbol{\sigma}|\boldsymbol{\beta}, \boldsymbol{\delta}, Y)$, and $f(\boldsymbol{\delta}|\boldsymbol{\beta}, \boldsymbol{\sigma}, Y)$. One parameter is drawn from each distribution in sequence, conditional on the most recently

sampled values of the other parameters, and the data $Y$. Sampling $\boldsymbol{\delta}$ consists of sub-steps in which each element $\delta_i$ is sampled conditional on the remaining elements $\boldsymbol{\delta}_{(-i)}$. While this may slow the convergence of the algorithm, it greatly simplifies the task of sampling from conditional distributions. The draw for $\boldsymbol{\beta}$ is a multivariate normal, $\sigma^2$ is an inverse gamma draw, and each individual $\delta_i$ is a Bernoulli draw. Further details of this procedure, its convergence and the exact form of the conditional distributions used may be found in GM.

What is of interest here is the choice of prior for the vector $\boldsymbol{\delta}$. Although their theory is general, in practice GM use an independence prior:

$$f(\boldsymbol{\delta}) = \prod_{i=1}^{p} p_i^{\delta_i}(1-p_i)^{1-\delta_i}, \qquad (3)$$

where $p_i = \Pr(\delta_i = 1)$. That is, the importance of any variable is independent of the importance of any other variable. This is a very parsimonious representation of prior knowledge, and in many situations it is quite accurate and appropriate. However, as the interaction example in the introduction illustrates, independence is not always appropriate.

## 3 Priors for Related Predictors

In this section, priors for $\boldsymbol{\delta}$ that incorporate relations between predictors are developed. The assumptions made in this development may be viewed as qualitative representations of commonly utilized principles of variable selection. Although the $\boldsymbol{\delta}$ notation used originates in the GM approach, the priors are applicable to any linear model and Bayesian variable selection technique.

### 3.1 Relations for Two-Way Interactions

Consider a simple example in which there are three main effects A, B, C and three two-way interactions AB, AC, and BC. The goal is to formulate a prior for $\boldsymbol{\delta}$ that allows the importance of interactions to depend on the importance of their parents.

After factoring, the joint density of $\boldsymbol{\delta}$ can be simplified by assuming that $\delta_A, \delta_B, \delta_C$ are independent, and that conditional on $(\delta_A, \delta_B, \delta_C)$, the interactions $(\delta_{AB}, \delta_{AC}, \delta_{BC})$ are independent:

$$\begin{aligned}\Pr(\boldsymbol{\delta}) &= \Pr(\delta_A, \delta_B, \delta_C, \delta_{AB}, \delta_{AC}, \delta_{BC}) \\ &= \Pr(\delta_A, \delta_B, \delta_C)\Pr(\delta_{AB}, \delta_{AC}, \delta_{BC}|\delta_A, \delta_B, \delta_C) \qquad (4)\\ &= \Pr(\delta_A)\Pr(\delta_B)\Pr(\delta_C)\Pr(\delta_{AB}|\delta_A, \delta_B, \delta_C)\Pr(\delta_{AC}|\delta_A, \delta_B, \delta_C)\Pr(\delta_{BC}|\delta_A, \delta_B, \delta_C)\end{aligned}$$

This expression can be further simplified by assuming that the importance of an interaction depends only on the importance of those main effects from which it was formed. That is, $\delta_{AB}$ depends on $\delta_A$ and $\delta_B$, but not $\delta_C$:

$$\Pr(\boldsymbol{\delta}) = \Pr(\delta_A)\Pr(\delta_B)\Pr(\delta_C)\Pr(\delta_{AB}|\delta_A, \delta_B)\Pr(\delta_{AC}|\delta_A, \delta_C)\Pr(\delta_{BC}|\delta_B, \delta_C). \qquad (5)$$

These two assumptions simplify the structure of the joint density for $\boldsymbol{\delta}$ greatly. Their importance is not simply a matter of mathematical convenience; these assumptions correspond to principles for variable selection that are commonly utilized. Both principles can be related to the hierarchical nature of the different model terms, in which the main effects are of the lowest (or simplest) order, and two-way interactions are of a higher (and thus more complex) order. The assumption that terms of a given order are independent, conditional on all terms of a lower order will be called the *conditional independence principle*. The second assumption, which states that a higher order term depends only on the lower order terms that were used to form it, will be called the *inheritance principle*.

The exact nature of this dependence on "parent" terms is defined by the components of the joint probability in (5). For example, the probability that the term $AB$ is active $\Pr(\delta_{AB} = 1|\delta_A, \delta_B)$ may take on four different values, depending on the values of the pair $(\delta_A, \delta_B)$:

$$P(\delta_{AB} = 1|\delta_A, \delta_B) = \begin{cases} p_{00} & \text{if } (\delta_A, \delta_B) = (0,0) \\ p_{01} & \text{if } (\delta_A, \delta_B) = (0,1) \\ p_{10} & \text{if } (\delta_A, \delta_B) = (1,0) \\ p_{11} & \text{if } (\delta_A, \delta_B) = (1,1) \end{cases}. \tag{6}$$

The choice of these values may represent different principles of variables selection. For example one might choose $(p_{00}, p_{01}, p_{10}, p_{11}) = (0, 0, 0, p)$. This would correspond to the prior belief that for an interaction to be active, all (in this case both) corresponding main effects must also be active. This principle has several names, including the "marginality assumption" (McCullagh and Nelder, 1989), and "well formulated models" (Peixoto, 1987, 1990). A less restrictive choice of the conditional probabilities might be $(p_{00}, p_{01}, p_{10}, p_{11}) = (0, p_1, p_2, p_3)$, corresponding to the belief that an interaction may enter the model if one or more corresponding main effect is in the model. Hamada and Wu (1992) call this the "effect heredity" principle. Friedman's (1991) MARS also utilizes a similar principle in its stepwise selection algorithm, since it may include an "interaction" between predictor A and B only if one of A or B is already in the model. A new naming convention will be adopted here, with the *strong heredity principle* corresponding to $(0, 0, 0, p)$ and the *weak heredity principle* corresponding to $(0, p_1, p_2, p_3)$.

In both cases, the implicit ordering $p_{00} \leq (p_{01}, p_{10}) \leq p_{11}$ seems natural. When there is little prior knowledge the same four probabilities could be used for all interaction terms. Prior knowledge that is specific to certain variables or collections of variables, could also be used in (5). The magnitude of the values chosen may be related to another variable selection principle, namely "effect sparsity", a term coined by Box and Meyer (1986). By making the prior probability of an active predictor small, the principle of the "magnificent few and the trivial many" predictors is expressed. This principle will be called the *sparsity principle*. Note that it may be applied to both main effects and higher order terms.

It is clear that by setting one or several of $p_{00}, p_{01}, p_{10}$ to be exactly zero, the model space is reduced. The number of models with positive mass can be calculated for varying rules; here calculations for strong and weak heredity principles are presented. Suppose there are $m$ main effects, and all possible two-way interactions are considered in the model. Then there are $p = m + \binom{m}{2}$ candidate terms, making a total of $2^p$ models. If strong heredity is assumed, (i.e. $p_{00} = p_{01} = p_{10} = 0$), then there are

$$\sum_{i=0}^{m} \binom{m}{i} 2^{\binom{i}{2}} \qquad (7)$$

models with nonzero mass. In the more relaxed case of weak heredity, (i.e. $p_{00} = 0$, others $> 0$), there are

$$\sum_{i=0}^{m} \binom{m}{i} 2^{mi - i(i+1)/2} \qquad (8)$$

models with nonzero mass. Proofs of (7) and (8) are given in the appendix.

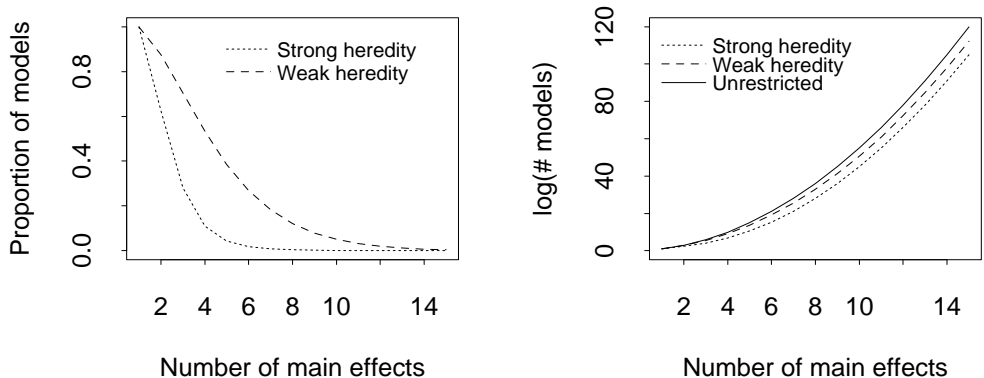

Figure 1: Number of models for restrictions on interactions given parents

In Figure 1, the total number of models under these two assumptions is compared to $2^p$, the total number of models. Two representations of the size of the model space are given in this figure. In the first, the proportion of the model space with nonzero prior probability is plotted. As the number of main effects increases, a smaller percentage of the whole model space has positive mass, and the more restrictive strong heredity principle (the finely dotted line) reduces the percentage most rapidly. In the second plot, the log of the number of models is plotted against the number of main effects. A base two log is used so that the vertical scale is an "effective number of predictors", since there are $2^p$ models without any restriction. Although the proportion of allowed models decreases with the number of main effects, there are still a very large number of models in all three cases. Other calculations could be presented for more complicated relations or for subsets of all possible interactions, but this gives a flavor of these restrictions.

When is the hierarchical structure useful? Although it can depend on the context, several general comments may be made. If the model is an approximation, first order terms are linear approximations while higher order terms are nonlinear, complex, and less stable. The hierarchical idea of including lower order terms first gives simpler and more stable models.

Weak heredity is useful because of its flexibility. Unlike weak heredity, strong heredity is invariant to linear transformations of predictors. A proof is given in Peixoto (1990); the following illustrates that weak heredity is not invariant. Consider the variables $A$, $B$, and $AB$, and suppose a model $\{A, AB\}$ is selected. Suppose the transformed variables $C = A+1$, $D = (B+1)/3$ are created and $\{C, CD\}$ is selected. This model involves all of the original terms, a different model from $\{A, AB\}$.

One case in which strict hierarchical structure is not useful is models that have active higher order terms without active lower order terms. For example, in the atmospheric sciences, relations of the form $Y = A \exp(BC)$ occur (see Kraght 1976). A log transform yields $\log(Y) = \log(A) + BC$, a linear model containing an interaction with no parents. Hierarchical relations can be made applicable by relaxing the priors. Instead of setting any of the probabilities $p_{00}, p_{01}, p_{10}, p_{11}$ to be exactly zero, some small number $\varepsilon$ is used. Strong heredity $(0, 0, 0, p_{11})$ is thus relaxed to $(0, \varepsilon, \varepsilon, p_{11})$ or $(\varepsilon, \varepsilon, \varepsilon, p_{11})$, and weak heredity $(0, p_{01}, p_{10}, p_{11})$ relaxed to $(\varepsilon, p_{01}, p_{10}, p_{11})$. When the data presents overwhelming evidence contrary to heredity, the posteriors for $\boldsymbol{\delta}$ will reflect this. The term "relaxed weak (strong) heredity" will refer to the relaxed form of weak (strong) heredity. With relaxed heredity priors, fewer models will be impossible, and the concentration of prior mass on the model space that will be of interest.

## 3.2 Hierarchical Polynomial Interactions

The principles and methods for expressing priors used in the previous section may be generalized to the case of models containing interactions between an arbitrary number of terms, each of an arbitrary order. Consider first a simple example, with the terms $A^2B^2$, $AB^2$, $A^2B$, $A^2$, $AB$, $B^2$, $A$, and $B$. The order of a term is defined as the total exponent of all components of each term. Thus $AB^2$ and $A^2B$ are of the same order, and of a lower order than $A^2B^2$. A term is said to *inherit* from a collection of lower order terms if it is equal to the product of those lower order terms. Thus, $A^2B$ inherits from $A^2, AB, A, B$ but not $B^2$. These relations can be expressed using a directed graph, as in Figure 2. As in the previous section, the density for $\boldsymbol{\delta}$ may be factored into a chain of terms of a given order, each conditional on all terms of a lower order. The conditional independence principle states that all terms of a given order are independent, conditional on all lower order terms. The inheritance principle is then used to state that a term depends only on those terms from which it inherits.

A third principle may achieve further simplification. A term *inherits immediately* from another term if it inherits from that term, and the term is of the next lowest order. That is $AB^2$ inherits immediately from $AB$ and $B^2$, and it inherits (but not immediately) from $A$ and $B$. The terms "parent" and "child" will be used in the context of immediate inheritance. The *immediate inheritance principle* is then defined as the assumption that given the importance of its parents, the importance of a child term is independent of all other terms. Strong

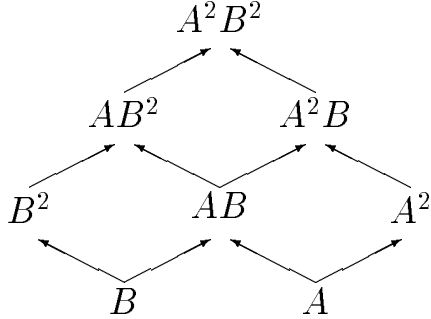

Figure 2: Ordering and inheritance relations among polynomial interactions

heredity would now require all parents to be active, and weak heredity would require only one.

Two interesting research topics related to these hierarchical structures are probability networks and image restoration. In the former, directed graphs are used to represent probability networks and the way in which information is propagated across them (see Lauritzen and Spiegelhalter, 1988). The networks and the form of the constructed densities are similar to those given here - conditioning and independence assumptions feature prominently. The main similarity is the concept of local dependence - the assumption that a node (here a variable like $\delta_A$) has several immediate neighbors which influence its probabilistic behavior, but that conditional on these neighbors, its distribution does not depend on more distant neighbors. This principle is also central in image restoration (see Geman and Geman, 1984), in which pixels are viewed as nodes.

The power of such an assumption is that it greatly simplifies the model, while still allowing a network of indirect links across all nodes $\delta_i$. The number of parents and children of a given node $\delta_i$ is greatly reduced by the immediate inheritance assumption, which may be viewed as a type of local structure. This local structure simplifies the computations necessary to implement this approach, since calculation of probabilities of the form $\Pr(\delta_i = \delta^* | \delta_{(-i)})$ up to a normalizing constant will take advantage of the factored nature of the prior.

The original and simplified forms of the prior on $\boldsymbol{\delta}$ may be expressed in general. Suppose that the highest order term is of order $k$. Let $\mathcal{O}_i$ represent all terms of a given order $i$, $\mathcal{F}(\delta_i)$ represent the family of all lower order terms from which $\delta_i$ inherits. Let $\mathcal{P}(\delta_i)$ represent the parents of the term $\delta_i$, namely those terms from which $\delta_i$ immediately inherits. Assume that there are $p$ components of $\boldsymbol{\delta}$. Then the density for $\boldsymbol{\delta}$ may be factored by the order of its components:

$$\Pr(\boldsymbol{\delta}) = \prod_{i=1}^{k} \Pr(\mathcal{O}_i | \mathcal{O}_j, j = 1, ..., i-1). \tag{9}$$

The conditional independence principle allows the densities for $\mathcal{O}_i$ to be broken into densities for individual elements $\delta' \in \mathcal{O}_i$,

$$\Pr(\boldsymbol{\delta}) = \prod_{i=1}^{k} \prod_{\delta' \in \mathcal{O}_i} \Pr(\delta' | \mathcal{O}_j, j = 1, ..., i - 1). \tag{10}$$

The inheritance principle reduces the set of terms upon which $\delta' \in \mathcal{O}_i$ depends from $\{\mathcal{O}_j, j = 1, ..., i - 1\}$ to the family of the term $\delta'$, denoted by $\mathcal{F}(\delta_i)$.

$$\Pr(\boldsymbol{\delta}) = \prod_{i=1}^{k} \prod_{\delta' \in \mathcal{O}_i} \Pr(\delta' | \mathcal{F}(\delta')) = \prod_{i=1}^{p} \Pr(\delta_i | \mathcal{F}(\delta_i)). \tag{11}$$

Finally, the immediate inheritance principle reduces the family $\mathcal{F}(\delta_i)$ to the set of parents $\mathcal{P}(\delta_i)$, giving

$$\Pr(\boldsymbol{\delta}) = \prod_{i=1}^{p} \Pr(\delta_i | \mathcal{P}(\delta_i)) = \prod_{i=1}^{p} p_i^{\delta_i} (1 - p_i)^{(1 - \delta_i)}, \tag{12}$$

where in the final expression, $p_i = \Pr(\delta_i = 1 | \mathcal{P}(\delta_i))$. This expression has structure similar to the independence prior originally considered in (3), with the crucial difference that the terms $p_i$ now are conditional probabilities.

While these relations seem like a sensible structure in the absence of exact contextual knowledge, others may be possible. For example, the immediate inheritance principle might not be appropriate, and instead it could be assumed that $\delta_i$ depends on its entire family $\mathcal{F}(\delta_i)$. Alternatively, the inheritance relations could be modified so that higher order terms depended on only main effect terms rather than terms of the next lowest order.

Relations for interactions that put zero (or little) mass on some models can influence the mixing behavior of stochastic search algorithms such as SSVS, or MC$^3$. Methods that traverse the model space by some sort of random walk will be more restricted when certain paths are not open to them, which would be caused by these priors.

### 3.3 Grouped Predictors

A type of predictor that commonly occurs in regression situations is a variable with many qualitative levels, such as treatment, supplier, or location. Since there is no ordered continuous scale for a multilevel qualitative factor, $l$ dummy variables are used (in the case of a predictor with $l + 1$ distinct levels). Quite often (in ANOVA, for example), the dummy variables are treated together, and are all included or excluded. This section incorporates this relation into the prior for $\boldsymbol{\delta}$.

Suppose for the purpose of illustration, that there is a group of variables $Q_1, Q_2, ..., Q_l$. The *grouping principle* reduces the set of possible values for $\boldsymbol{\delta}$. That is, only $(\delta_{Q_1}, \delta_{Q_2}, ..., \delta_{Q_l}) \in \{(0, 0, ..., 0), (1, 1, ..., 1)\}$ are considered. A simpler way to represent this restriction is to note that there is actually a single scalar $\delta_Q$ which determines the importance of $Q_1, Q_2, ..., Q_l$. That is, there is a many-to-one mapping from the vector of regression coefficients $\boldsymbol{\beta}$ to the vector of latent variables $\boldsymbol{\delta}$, as shown in Figure 3. The grouping principle need not be constrained to qualitative predictors. There may be other situations in which certain predictors belong to a larger class, and interest focuses on whether that class is influential.

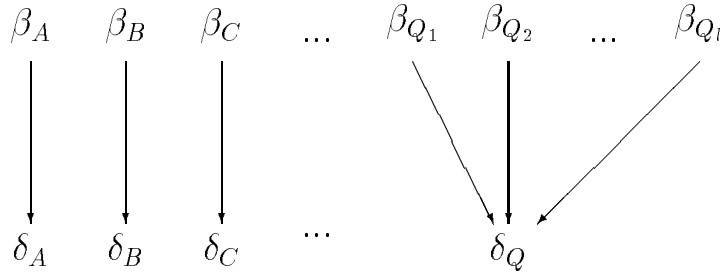

Figure 3: The many-to-one mapping of regression coefficients to latent variables

The hierarchical relations defined in the previous two subsections can be used in conjunction with the grouping principle. For example, if $A$ is some other variable, the $AQ$ interaction will consist of (at least) $l-1$ variables. A single element $\delta_{AQ}$ would indicate the importance of these dummy variable interactions, and heredity relations would describe the dependence of $\delta_{AQ}$ on $\delta_A$ and $\delta_Q$.

Suppose that $\Pr(\delta_Q = 1|Y)$ is large. The first question that comes to mind is whether only certain components of $Q$ are influential. Rather than refit the model, the posterior of the regression coefficients $\beta_{Q_1}, \beta_{Q_2}, ..., \beta_{Q_l}$ can answer this question. The event $\delta_Q = 1$ does not imply that all of $\beta_{Q_1}, \beta_{Q_2}, ..., \beta_{Q_l}$ are large, but only that some of them are. By examining plots of the joint posterior for $\beta_{Q_1}, \beta_{Q_2}, ..., \beta_{Q_l}$, it may be possible to identify directions in which effects are larger or smaller.

Not only does the grouping principle reduce the size of the total model space, but it makes headway in dealing with the pitfalls of multiple comparisons. When a group of variables is not active, but there are many of them, the chances are large that one variable will appear active because of random variation. By only considering the importance of the group, the chances of making such an error are reduced, because either a large single effect or several medium sized effects will be needed to conclude that the group is active. A striking example of this property is given in Section 4.2.

In terms of implementation, the prior for $\boldsymbol{\delta}$ is of the same form as before, but with fewer elements. The prior for $\boldsymbol{\beta}$ is the component influenced by this modification, since $\beta_{Q_1}, \beta_{Q_2}, ..., \beta_{Q_l}$ will all have variances that depend on $\delta_Q$.

### 3.4 Competing Predictors

Another relation between predictors is a competitive one, in which either one predictor or the other is active, but not both. This sort of relation could occur if competing scientific hypotheses are being tested, or transformations of predictor variables are to be chosen from a small set of candidates. The latter case would involve a choice between (say) $A$, $\sqrt{A}$, $\log A$ or $A^{-1}$ as predictors. Unless the model is quite complex, only one of these predictors would be included in the model.

Consider the following illustrative example. Suppose that predictors $A$, $B$, and $C$ are observed, and all two-way interactions between these three factors are considered. Values of another variable $a$ are available, but models with both $A$ and $a$ make no sense. The variables $a$ and $A$ are competing predictors, as are the two-way interactions $\{aB, aC\}$ and $\{AB, AC\}$. The model space may be divided into two subspaces: $\{A, B, C, AB, AC, BC\}$, and $\{a, B, C, aB, aC, BC\}$. Using the notation $M_i$ and $\overline{M}_i$ to represent one of the subspaces and its complement,

$$M_1 = A, B, C, AB, AC, BC \qquad \overline{M}_1 = a, aB, aC \qquad (13)$$
$$M_2 = a, B, C, aB, aC, BC \qquad \overline{M}_2 = A, AB, AC.$$

Note that the competing predictors restriction requires that if $\boldsymbol{\delta}_{M_i} \neq 0$ then $\boldsymbol{\delta}_{\overline{M}_i} = 0$. The prior for $\boldsymbol{\delta}$ may be written as a $(p_1, 1 - p_1)$ mixture of two models, one with terms not involving $a$, and the other not involving $A$.

$$\Pr(\boldsymbol{\delta}) = \Pr(\boldsymbol{\delta}_{M_1}) I(\boldsymbol{\delta}_{\overline{M}_1} = 0) p_1 + \Pr(\boldsymbol{\delta}_{M_2}) I(\boldsymbol{\delta}_{\overline{M}_2} = 0)(1 - p_1) \qquad (14)$$

More generally, if there are $k$ models, the mixture would be

$$\Pr(\boldsymbol{\delta}) = \sum_{i=1}^{k} \Pr(\boldsymbol{\delta}_{M_i}) I(\boldsymbol{\delta}_{\overline{M}_i} = 0) p_i. \qquad (15)$$

with mixing probabilities $p_1, \ldots, p_k$ such that $\sum_{i=1}^{k} p_i = 1$. Assuming the principles outlined in the previous section hold, the terms in (14) may be simplified. For example, the conditional independence principle and the inheritance principle imply (5). Once the same simplification is made for $\Pr(M_2)$, the prior on $\boldsymbol{\delta}$ in (14) may be expressed as

$$\Pr(\boldsymbol{\delta}) = \Pr(\delta_B) \Pr(\delta_C) \Pr(\delta_{BC}|\delta_B, \delta_C) \times$$
$$\{\Pr(\delta_A) \Pr(\delta_{AB}|\delta_A, \delta_B) \Pr(\delta_{AC}|\delta_A, \delta_C) I(\{\delta_a, \delta_{aB}, \delta_{aC}\} = \mathbf{0}) p_1 + \qquad (16)$$
$$\Pr(\delta_a) \Pr(\delta_{aB}|\delta_a, \delta_B) \Pr(\delta_{aC}|\delta_a, \delta_C) I(\{\delta_A, \delta_{AB}, \delta_C\} = \mathbf{0})(1 - p_1)\}$$

Note that in (16) terms which are common to both models (here $B, C, BC$) may be factored out of the individual components of the equation. In this case, the conditional independence principle and the heredity principle allowed the terms to be factored out. This is generally the case, since conditional independence breaks up the probability into a product of terms, and inheritance removes any competing terms from probabilities for terms common to all models.

The observations regarding mixing of stochastic search algorithms made in the section on polynomial relations apply here, since competing predictor priors assign zero mass to a large portion of the model space. Specifically, all paths between two competing models $M_1$ and $M_2$ would be through their intersection, in which no competing terms from either model is active.

### 3.5 Global Constraints on Models

This section considers a supplementary method for writing out the prior on $\boldsymbol{\delta}$ which is based on more global constraints. In general, suppose that some probability model (i.e., a prior) for $\boldsymbol{\delta}$, denoted by $f(\boldsymbol{\delta})$ has already been specified. This prior could be weighted according to some global property of $\boldsymbol{\delta}$, using a weighting function $w(\boldsymbol{\delta})$:

$$f_w(\boldsymbol{\delta}) \propto w(\boldsymbol{\delta})f(\boldsymbol{\delta}). \tag{17}$$

The most obvious global property that $w(\boldsymbol{\delta})$ might represent is the number of active terms in the model, and this was the original motivation for using such a construction. For example, $w(\boldsymbol{\delta}) = I(|\boldsymbol{\delta}| \leq 10)$ where the number of active terms is given by $|\boldsymbol{\delta}| = \sum_{i=1}^{p} \delta_i$ would restrict attention to those models with fewer than ten active terms. GM suggest the prior $f_w(\boldsymbol{\delta}) = w_{|\boldsymbol{\delta}|}\binom{p}{|\boldsymbol{\delta}|}^{-1}$ where $w_{|\boldsymbol{\delta}|} = \Pr(|\boldsymbol{\delta}|$ terms active). This is another example of weighting with $f(\boldsymbol{\delta}) =$ constant. Variations on the number of terms active could include the number of terms of a given order.

No matter how it is defined, the weight $w(\boldsymbol{\delta})$ will be discretely valued since $\boldsymbol{\delta}$ has a finite number of elements. This means that $w(\boldsymbol{\delta})$ may be viewed as defining a partition of the model space, and then reweighting each partition. Interest will focus on simple partitions such as the examples given above. In general the weighted partition will not preserve the probability structure described in the previous sections. Consider a simple case involving main effects A, B, and C. Assume independence of $\delta_A, \delta_B$, and $\delta_C$, and reweight $\boldsymbol{\delta} = (1, 1, 1)$ to have zero prior mass. Then $\Pr(\delta_A = 1 | \delta_B = \delta_C = 1) = 0$ but $\Pr(\delta_A = 1) \neq 0$. Note that conditional on any given partition defined by $w$, the structures and relations defined previously still hold. Also, it is clear that any strong restrictions (those which assign probability zero to certain models, such as strong heredity) will still hold. What will change is the probabilities of certain events (due to the reweighting), and independence assumptions. The extent of this change will depend on the unweighted probability of the partitions and the weighting function.

If the priors are constructed properly using the suggestions in the previous sections, global restrictions may be unnecessary. It is recommended that the unweighted prior probabilities of each proposed partition be determined (either analytically or through some sampling approach) *before* any partitions are reweighted. Even when this is done, it may be believed that certain partitions are in fact impossible, and reweighting may be justified.

This prior is the easiest to implement, since the reweighting function can simply be multiplied by the prior after the latter is calculated. Typically this involves calculation of the number of active terms.

## 4 Constructed Examples

In this section, data are generated from a number of known models, and a comparison of model selection made between an independence prior and the appropriate related predictor prior. A small factorial experiment is conducted, with three factors: the true model, the amount of noise, and the prior used for variable selection.

| Description | $\Pr(\delta_A = 1)$ | $\Pr(\delta_{AB} = 1 \mid \delta_A, \delta_B)$ | model | error |
|---|---|---|---|---|
| strong heredity | 0.5 | $(0.00, 0.00, 0.00, 0.50)$ | $\boldsymbol{\beta} = (1, 1, 0, 1, 0, 0)$ | $\sigma = 1$ |
| relaxed strong heredity | 0.5 | $(0.00, 0.01, 0.01, 0.50)$ | $\boldsymbol{\beta} = (1, 0, 0, 1, 0, 0)$ | $\sigma = 3$ |
| weak heredity | 0.5 | $(0.00, 0.25, 0.25, 0.50)$ | $\boldsymbol{\beta} = (0, 0, 0, 1, 0, 0)$ | |
| relaxed weak heredity | 0.5 | $(0.01, 0.25, 0.25, 0.50)$ | | |
| iid | 0.5 | $(0.50, 0.50, 0.50, 0.50)$ | | |
| independent | 0.5 | $(0.25, 0.25, 0.25, 0.25)$ | | |

Table 1: Levels for factorial experiment, interaction example

| | $\sigma = 1$ | | | $\sigma = 3$ | | |
|---|---|---|---|---|---|---|
| Prior | $A, B, AB$ | $A, AB$ | $AB$ | $A, B, AB$ | $A, AB$ | $AB$ |
| strong heredity | 0.966 | 0.000 | 0.000 | 0.885 | 0.000 | 0.000 |
| relaxed strong heredity | 0.964 | 0.128 | 0.000 | 0.799 | 0.045 | 0.000 |
| weak heredity | 0.858 | 0.741 | 0.000 | 0.381 | 0.419 | 0.000 |
| relaxed weak heredity | 0.866 | 0.750 | 0.133 | 0.355 | 0.419 | 0.048 |
| iid | 0.677 | 0.602 | 0.545 | 0.153 | 0.205 | 0.335 |
| independent | 0.816 | 0.742 | 0.648 | 0.223 | 0.250 | 0.452 |

Table 2: Posterior probability of true model, interaction example

### 4.1 Inheritance

This example involves a sample of 50 observations, with terms $A$, $B$, $C$, $AB$, $AC$, $BC$ as possible predictors. Six different priors are compared under three different models, each at two levels of noise. The levels used for these three variables are given in Table 1. The six priors represent strong heredity, weak heredity, and independence. For each of the heredity priors, a "relaxed" version with 0.01 replacing 0 is considered, and two different independence priors are used. The data are generated from three models; the first contains $A$,$B$,$AB$, the second $A$, $AB$, and the last only $AB$. Two levels of noise are considered, giving $6 \times 3 \times 2 = 36$ possible combinations. In order to facilitate comparisons, the same levels of $A$, $B$, $C$ and a single $\epsilon$ vector of 50 $N(0,1)$ errors multiplied by one or three are used for all 36 experiments. The values of $A$, $B$, and $C$ are generated as iid $N(0,1)$. Values of $\tau = 0.2, c = 10$ are used. In each of the 36 cases, 1000 samples from the posterior were realized by taking every 10th sample from a single long run of 10000 samples.

The result of each experiment is an approximate sample from the posterior on the $2^6 = 32$ possible models. This outcome is summarized in Table 2, which gives the posterior probability of the correct model under each of the experimental settings. As one would expect, strong heredity does best under model one, weak under model two, and independence under the third model. In cases where the principles of a prior are violated (e.g strong heredity under models two and three), no mass is put on the correct model. Some balance may be achieved by relaxing the strict hierarchical assumptions. When $\sigma = 1$, some posterior mass is placed on the correct model (0.128 for $A, AB$ under relaxed strong heredity and 0.133 for $AB$ under relaxed weak heredity). Although not large, it represents a large shift from the prior, which assigns a probability of 0.0012 on the true model in both cases. This large change in prior mass, which can be identified by posterior to prior odds, should be enough to draw attention to the true model in both cases.

| Description | Pr($\delta_A = 1$) | Pr($\delta_C = 1$) | model | error |
|---|---|---|---|---|
| grouping | 0.5 | 0.20 | $\boldsymbol{\beta} = (1, 0, 1, 1, 1, 1)$ | $\sigma = 1$ |
| grouping | 0.5 | 0.50 | $\boldsymbol{\beta} = (1, 0, 2, 0, 0, 0)$ | $\sigma = 3$ |
| grouping | 0.5 | 0.80 | $\boldsymbol{\beta} = (1, 0, 1, 0, 0, 0)$ | |
| iid | 0.5 | — | $\boldsymbol{\beta} = (1, 0, 0, 0, 0, 0)$ | |

Table 3: Levels for factorial experiment, grouping example

As the noise increases, there is generally less certainty about which model is most probable. In all cases, the true model receives less mass, and there is no longer sufficient information in the data for relaxed versions of weak and strong heredity to suggest the true model as a possibility.

One other interesting feature of this experiment not illustrated in the table is the tendency of the posterior to "pull in" inactive parents when strong or weak heredity is assumed and an interaction is important. For example, under strong heredity, $\sigma = 1$, and a model with only $AB$ active, the model $A, B, AB$ is the most probable, with posterior mass 0.966, and a marginal probability of 1.000 that $AB$ is active (this is in part due to the large magnitude of the estimated $AB$ coefficient, but still indicative of the behavior of this method). When heredity assumptions are violated, unimportant terms are likely to be brought into the model, rather than important terms omitted. This behaviors mimics the "common practice" of adding main effects when corresponding interactions are important.

## 4.2 Grouping

This example compares grouping and independence priors. The candidate predictors are $A$, $B$ and a categorical variable $C$, which takes on five distinct levels, represented by four dummy variables $C_1$, $C_2$, $C_3$, and $C_4$. As in the previous example, the performance of priors under different true models is investigated via a factorial experiment, with the three "factors" being the prior, true model, and amount of noise. The levels of these are given in Table 3. The grouping priors use a single $\delta_C$ to indicate whether the four dummy variables for $C$ are active; the independence prior uses individual $\delta$'s for each of the four $C_i$'s. For all four priors, the terms $A$ and $B$ have prior probabilities of 0.5 of being active. The iid prior also gives each of $C_1, \ldots, C_4$ a probability of 0.5 of being active. Because $\delta_C$ represents four dummy variables, it is not clear whether it should have the same probability of being active. This possibility is explored through the use of three different grouping priors, which assign probabilities 0.2, 0.5, and 0.8 to the event $\delta_C = 1$. Values of $\boldsymbol{\tau} = (0.2, , 0.2, 0.08, 0.08, 0.08, 0.08), c = 10$ were used.

The experiment is carried out as before. The results are summarized in terms of the marginal probability that a term is active. The large number of models under the indepen-dence prior in which one or more of $C_1$ are included makes comparison of full models difficult. Table 4 gives the marginal probability that each term is active. The last two rows of the table give the marginal probability that under an independence prior, all of the $C_i$'s (first row) or any of the $C_i$'s (second row) are active. Such alternatives to formal grouping are considered by Clyde and Parmigiani (in press), with the basic idea being that a term is declared active based on the number of dummy variables included in the model. The table illustrates that requiring all terms to be active can result in false negatives (e.g. when $\sigma = 3$ and $C_1, \ldots, C_4$ are active, only 15.1% of the samples have all terms active), and requiring any can result in false positives (e.g. when all elements of $C$ are inactive, 80.4% and 92.1% of the samples have at least one term active).

|       | $\sigma = 1$ | | | | $\sigma = 3$ | | | |
| Prior | $A, C_1, \ldots, C_4$ | $A, 2C_1$ | $A, C_1$ | $A$ | $A, C_1, \ldots, C_4$ | $A, 2C_1$ | $A, C_1$ | $A$ |
| --- | --- | --- | --- | --- | --- | --- | --- | --- |
| group 0.20 | 1.000 | 1.000 | 0.923 | 0.011 | 0.985 | 0.551 | 0.277 | 0.150 |
| group 0.50 | 1.000 | 1.000 | 0.955 | 0.071 | 1.000 | 0.814 | 0.583 | 0.289 |
| group 0.80 | 1.000 | 1.000 | 1.000 | 0.217 | 1.000 | 0.998 | 0.860 | 0.633 |
| iid $\delta_{C1}$ | 0.994 | 1.000 | 0.992 | 0.256 | 0.685 | 0.842 | 0.581 | 0.408 |
| $\delta_{C2}$ | 0.844 | 0.299 | 0.310 | 0.304 | 0.492 | 0.449 | 0.452 | 0.480 |
| $\delta_{C3}$ | 1.000 | 0.585 | 0.557 | 0.527 | 0.912 | 0.670 | 0.681 | 0.571 |
| $\delta_{C4}$ | 0.977 | 0.235 | 0.273 | 0.229 | 0.549 | 0.363 | 0.416 | 0.413 |
| iid - all | 0.818 | 0.046 | 0.048 | 0.009 | 0.151 | 0.092 | 0.069 | 0.046 |
| iid - any | 1.000 | 1.000 | 0.999 | 0.804 | 0.995 | 0.979 | 0.955 | 0.921 |

Table 4: Marginal probability that dummy variables are active, grouping example

When there isn't much noise, the conclusions delivered by grouping are generally more decisive. This is most evident when all the dummy variables are inactive. Even with a prior probability of 0.8 that $\delta_C = 1$, the posterior based on grouping offers less evidence that $C$ is active than the iid prior.

When there is more noise, there is still some advantage to using grouping. For example, when $C$ is inactive, all but the 0.8/grouping prior give pretty strong support for the truth, whereas the iid prior suggests that some components of $C$ may be active. In cases where only one dummy variable is active, neither method gives a conclusive result, but the grouping does no worse.

It may seem that information is being lost under grouping if only some components of a categorical factor are important - for example, with $\sigma = 1$ and only $C_1$ active, the iid posterior indicates that only $C_1$ is active, unlike the grouping prior. However, additional information about the composition of the effect of $C$ is available from the posterior for $\boldsymbol{\beta}$, as discussed in Section 4.2.

| Code | Variable Description |
|------|----------------------|
| Y | Estimated performance measure |
| A | cycle time in nanoseconds |
| B | minimum main memory in kilobytes |
| C | maximum main memory in kilobytes |
| D | cache size in kilobytes |
| E | minimum number of channels |
| F | maximum number of channels |

Table 5: Variable names and descriptions, CPU data

## 5 CPU Performance Data

This section considers data on performance and characteristics of 209 computer central processing units (cpus). The data, which are available in Venables and Ripley (1994), are described in Table 5. For notational convenience the six predictor variables are labeled $A - F$. Each cpu has an estimated performance measure and six (continuous) characteristics. A full second order model with performance as the response is considered, with a total of 27 terms (6 main effects, 6 quadratic terms, and 15 two-way interactions). Quadratic terms and interactions are formed from the original variables without transformation. Collinearity among the 27 terms is strong, with correlations ranging from -0.38 to 0.97.

Why might someone be interested in a second order model? Two possible answers are: (1) There could be suspicion that the response surface is nonlinear and involves interactions. (2) There might be interest in fitting a nonparametric smooth surface including "interaction" terms of the form $f_{ij}(X_i, X_j)$. In (1), variable selection with these priors makes it possible to search for second order models that are simpler than the full 27 term model, and obey principles discussed in Section 3. In (2), there are too many possible interactions to fit nonparametrically. The proposed procedure will find a smaller set of interactions which may be used as a starting point for nonparametric models.

Values used for $\tau_i$ are $\tau_i = 6\hat{\sigma}_i$ where $\hat{\sigma}_i$ is the standard error of the least squares estimate $\hat{\beta}_i$ from a full regression on 27 predictors. As recommended in GM, different multiples of $\tau$ are tried; here the multiplier 6 is used for illustrative purposes. An uninformative prior on $\sigma$, and $c_i = 10$ are used. Three different priors on the models space are considered: strong heredity, relaxed weak heredity, and independence. The three priors are:

$$\boxed{\text{Strong Heredity}}: \text{Dependent, with inheritance: } \Pr(\delta_A = 1) = 0.50, \tag{18}$$
$$\Pr(\delta_{A^2} = 1|\delta_A) = (0, 0.50), \Pr(\delta_{AB} = 1|\delta_A, \delta_B) = (0, 0, 0, 0.25).$$

$$\boxed{\text{(Relaxed) Weak Heredity}}: \text{Dependent, with inheritance: } \Pr(\delta_A = 1) = 0.50, \tag{19}$$
$$\Pr(\delta_{A^2} = 1|\delta_A) = (0.05, 0.50), \Pr(\delta_{AB} = 1|\delta_A, \delta_B) = (0.05, 0.10, 0.10, 0.25).$$

$$\boxed{\text{Independence}}: \text{Independent, terms of same type identically distributed,} \tag{20}$$
$$\Pr(\delta_i = 1) = \{0.50, 0.50, 0.10\} \text{ for main effects, quadratics, and interactions.}$$

Recall that the notation $\Pr(\delta_{A^2} = 1|\delta_A) = (p_1, p_2)$ and $\Pr(\delta_{AB} = 1|\delta_A, \delta_B) = (p_1, p_2, p_3, p_4)$ refers to

$$\Pr(\delta_{A^2} = 1|\delta_A) = \begin{cases} p_1 & \text{if } \delta_A = 0 \\ p_2 & \text{if } \delta_A = 1 \end{cases}$$

and

$$\Pr(\delta_{AB} = 1|\delta_A, \delta_B) = \begin{cases} p_1 & \text{if } (\delta_A, \delta_B) = (0,0) \\ p_2 & \text{if } (\delta_A, \delta_B) = (0,1) \\ p_3 & \text{if } (\delta_A, \delta_B) = (1,0) \\ p_4 & \text{if } (\delta_A, \delta_B) = (1,1) \end{cases}.$$

The three priors have roughly the same marginal distribution on the number of active terms. Models under strong heredity are invariant to linear transforms of the original six predictors. Relaxed weak heredity allows more flexibility in the search, while focusing attention on models that are hierarchical. The large number of interactions in the model motivates the decision to prejudice priors against higher order terms through the use of smaller probabilities and hierarchical structures.

For each of the three priors, approximate posterior samples of size 10,000 are generated by taking every 50th value from a single chain of length 500,000. Autocorrelations are negligible in the sampled values, and repeated runs indicate that accuracy in probabilities for $\delta$ is roughly ±0.01. Smaller runs would in fact be sufficient for exploratory purposes. The SSVS algorithm takes 60 minutes on an Sun Sparcstation 20 for each posterior.

Table 6 summarizes the three posteriors. Only factors with a marginal probability of 0.25 or more in at least one posterior are included in the marginal table. Terms $B$, $C$, $D$, $C^2$, $D^2$, $BC$, $CD$ and $CF$ appear active in all three posteriors. The term $F$ has high probability of being active only in the strong heredity posterior, indicating that it is likely "drawn in" by the active $CF$ interaction. Strong and weak heredity posteriors have more mass on the top five models, due to their more concentrated priors. In all posteriors there is strong evidence of interactions and nonlinear behaviour.

In this example, two advantages that the hierarchical priors have over an independence prior are more decisive conclusions, and prejudice against higher order terms. The more concentrated posterior under heredity means that a few models stand out as likely, and other less attractive ones (which may not obey strong or weak heredity) have been filtered out. The prejudice against higher order terms may be seen by calculating Pr(no interactions active). With strong heredity, this is a priori 0.45; with an independence prior, it is 0.20. The concentration of prior mass on models without interactions reduces the chance of detecting spurious interactions.

Figure 4 gives a graphic representation of the posterior on $\boldsymbol{\delta}$. This plot may be thought of as a stretched matrix of 0's (white) and 1's (black). Each row of the matrix is a $\boldsymbol{\delta}$ from the posterior; the corresponding variables are labeled at the bottom. The $\boldsymbol{\delta}$'s in the plot are ordered from the most probable model (bottom) to the least (top). The ten most probable models are separated by horizontal lines to aid comparison. The vertical distance between lines is the posterior probability associated with that model. So, the second most probable

**Marginal Probabilities**

|  | B | C | D | E | F | C² | D² | BC | BF | CD | CF |
|---|---|---|---|---|---|---|---|---|---|---|---|
| strong | 0.995 | 1 | 1.000 | 0.105 | 0.993 | 1 | 0.980 | 0.947 | 0.616 | 1 | 0.975 |
| weak | 0.821 | 1 | 1.000 | 0.140 | 0.440 | 1 | 0.971 | 0.967 | 0.387 | 1 | 0.985 |
| indep. | 0.766 | 1 | 0.997 | 0.302 | 0.324 | 1 | 0.969 | 0.928 | 0.312 | 1 | 0.975 |

**Joint Probabilities (top 5 models for each posterior)**

|  | B | C | D | E | F | C² | D² | F² | BC | BF | CD | CF | DF | p | R² |
|---|---|---|---|---|---|---|---|---|---|---|---|---|---|---|---|
| Strong | • | • | • |  | • | • | • |  | • | • | • | • |  | 0.279 | 0.9988 |
| Strong | • | • | • |  | • | • | • |  | • |  | • | • |  | 0.189 | 0.9986 |
| Strong | • | • | • |  | • | • | • | • | • | • | • | • |  | 0.085 | 0.9988 |
| Strong | • | • | • |  | • | • | • | • | • |  | • | • |  | 0.050 | 0.9986 |
| Strong | • | • | • |  | • | • | • |  | • | • | • | • | • | 0.032 | 0.9988 |
|  |  |  |  |  |  |  |  |  |  |  |  |  | total | 0.635 |  |
| Weak | • | • | • |  |  | • | • |  | • |  | • | • |  | 0.172 | 0.9985 |
| Weak | • | • | • |  | • | • | • |  | • |  | • | • |  | 0.078 | 0.9988 |
| Weak | • | • | • |  | • | • | • |  | • |  | • | • |  | 0.053 | 0.9986 |
| Weak | • | • | • |  |  | • | • |  | • | • | • | • |  | 0.053 | 0.9986 |
| Weak |  | • | • |  |  | • | • |  | • |  | • | • |  | 0.049 | 0.9982 |
|  |  |  |  |  |  |  |  |  |  |  |  |  | total | 0.405 |  |
| Indep. | • | • | • |  |  | • | • |  | • |  | • | • |  | 0.096 | 0.9985 |
| Indep. | • | • | • | • | • | • | • |  | • |  | • | • |  | 0.034 | 0.9986 |
| Indep. | • | • | • | • |  | • | • |  | • |  | • | • |  | 0.031 | 0.9986 |
| Indep. | • | • | • |  |  | • | • |  | • | • | • | • |  | 0.029 | 0.9986 |
| Indep. |  | • | • |  |  | • | • |  | • |  | • | • |  | 0.023 | 0.9982 |
|  |  |  |  |  |  |  |  |  |  |  |  |  | total | 0.213 |  |

Table 6: Posterior probabilities for marginals and models. In the lower table, a • represents an active term, and terms not listed are excluded from all models.

model has roughly mass 0.20, and active terms $B$, $C$, $D$, $F$, $C^2$, $D^2$, $BC$, $CD$, and $CF$. If all the white space were removed from the plot, and the black rectangles "fell" to the bottom of the plot, a histogram of the marginal probability that a term is active would result. Additional information about the joint distribution of $\delta$ is available from the plot. For example, the four most probable models all have the terms $B$, $C$, $D$, $F$, $C^2$, $D^2$, $D^2$, $BC$, $CD$, and $CF$, and the only difference is whether $F^2$ and $BF$ are included.

This plot is a modification of an earlier plot (Chipman (1994), Clyde and Parmigiani (in press)), in which the matrix of 0's and 1's is not stretched vertically. That plot is useful for comparing models if their posterior mass is not the main interest. It is also a useful diagnostic: if the $\delta$ values are plotted in the order they are sampled, it is possible to see if the algorithm is getting "stuck" in a certain model or neighborhood.

It is interesting to compare this approach to conventional variable selection methods, such as stepwise and all-subsets regression. Almost all models found by conventional searches do not obey strong heredity, and the models found by stepwise regression are sensitive to the starting point used. In Figure 5, the residual sum of squares (RSS) is plotted against the number of terms for various models. The line represents models found by stepwise; the points the ten best models in the strong heredity posterior. The latter contains several different models with lower RSS than models found by stepwise. This is impressive, since no model found by stepwise with fewer than 18 terms obeys strong heredity. SSVS with strong heredity finds models as good or better than stepwise even though it searches a smaller space.

There are few enough predictors that an exhaustive search may be performed on the data. The vast majority of these models do not obey strong heredity. For example, only one of the best 50 models with 10 terms obeys strong heredity. The smallest RSS for a five term strong heredity model is 44,150, the 914th best model of this size! Comparison of this value to those in Figure 5 indicates that the curve for stepwise is unrealistically low, and models with more than five terms merit consideration. It is also clear that even when an exhaustive search is possible, it can be tedious finding models that fit well and are attractive.

In this example, grouping priors could be used for indicator variables describing the 30 companies that manufactured the chips. Either all 29 terms would be included or not. Without grouping priors, variable selection in the space of $2^{27+29} = 2^{56}$ models would be difficult. Grouping priors are not explored here because a preliminary investigation indicated that the company type had little value in predicting performance.

A few conclusions may be reached regarding the analysis of this data set. It appears that nonlinearities and interactions are present, and if such models are entertained, significantly lower RSS values are achievable (the RSS for a main effects only model is 444,008, compared to values under 20,000 for models identified here.) Only one main effect, the minimum number of channels ($E$) appears unimportant, and several nonlinearities and interactions appear interesting. The hierarchical priors with SSVS have identified parsimonious second order models which could be used model performance, or be used as a starting point for smooth modeling of specific interactions and nonlinearities.

This example typifies several characteristics of variable selection procedures that utilize related predictor priors. By concentrating the prior on models that are "interesting", the level of noise in the posterior is reduced. More complex models containing higher order terms are most likely to receive posterior mass when they include (simpler) lower order terms. Stochastic search techniques are more thorough than stepwise methods, and applicable even when exhaustive searches are not possible (see Chipman 1994, for an example). In large (say fifty or more terms), where model uncertainty is large, and deterministic searches difficult or impossible, this approach provides a promising alternative.

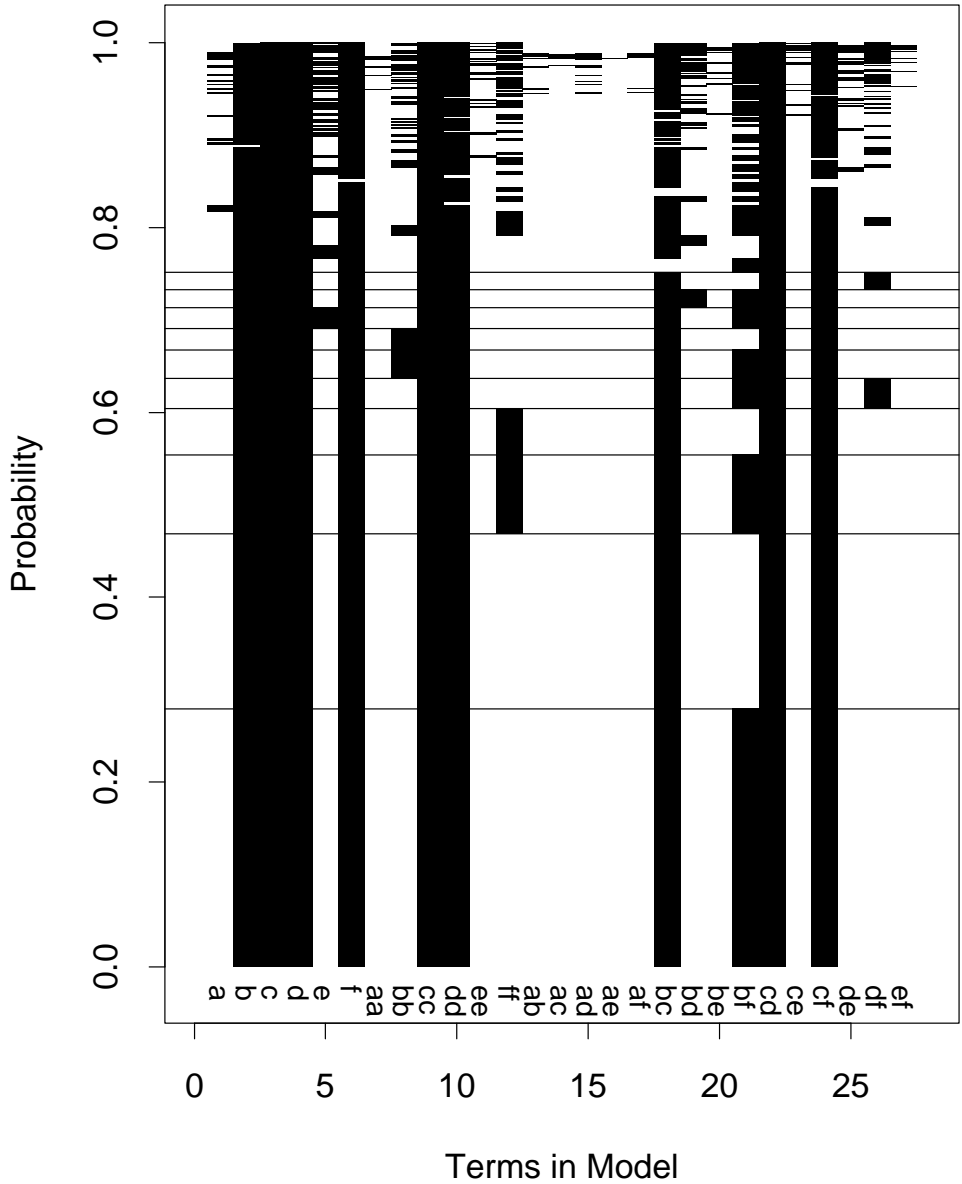

Figure 4: Graphic representation of the 100 most probable models, strong heredity posterior. Each "row" represents a model, and rows are stretched vertically in proportion to the posterior probability of the corresponding model. The ten most probable models are separated by horizontal lines.

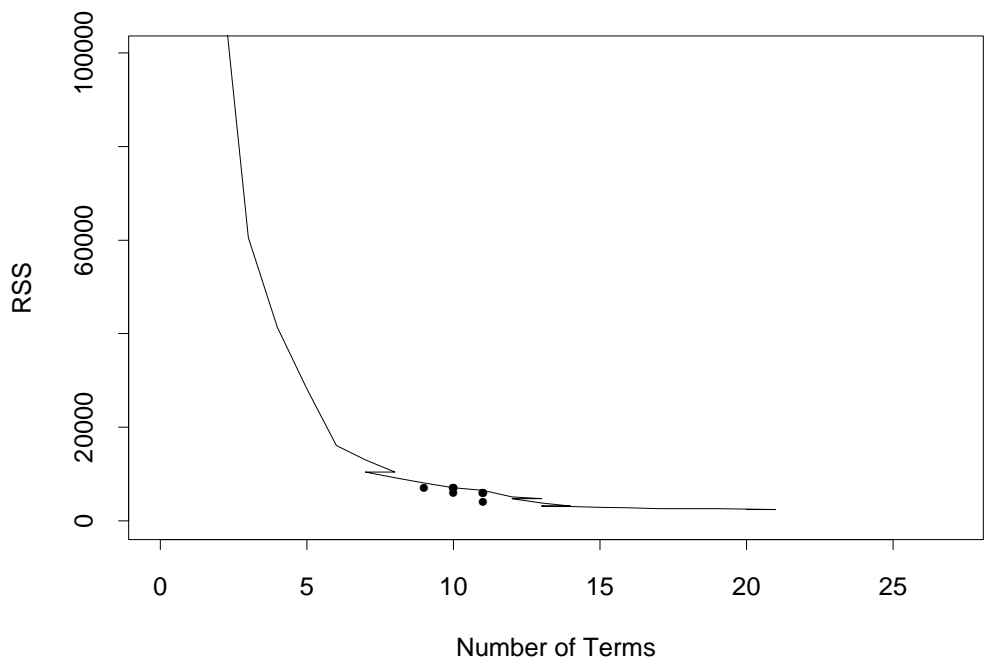

Figure 5: RSS for models found by stepwise (line) and the ten best found by SSVS with strong heredity (dots). Note that RSS without any predictors is 4,981,550.

# Appendix

### Proof of combinatorics for (7) and (8)

In both proofs, assume there are $m$ main effects and each of the $\binom{m}{2}$ two-way interactions between these effects are considered as possible predictors. For a set of active main effects, strong and weak heredity will require that certain interactions must be inactive, while others could be either active or inactive. This latter set will be referred to as interactions that are *eligible* to be active.

### Strong heredity (7):

The strong heredity principle states that an interaction is eligible to be active only when both main effect parents are active. For a set of $i$ active main effects there are $\binom{i}{2}$ interactions eligible to be active. Since each may be active or inactive independently of the others, there are a total of $2^{\binom{i}{2}}$ distinct models for each different set of $i$ active main effects. There are $\binom{m}{i}$ different sets of $i$ active main effects, so summing $i$ from 0 to $m$ yields a total of

$$\sum_{i=0}^{m} \binom{m}{i} 2^{\binom{i}{2}} \qquad (7)$$

distinct models under strong heredity.

### Weak heredity (8):

Consider a set of $i$ active main effects labeled $m_1, m_2, \ldots, m_i$. Under weak heredity, there are $m-1$ eligible two-way interactions involving $m_1$. An additional $m-2$ uncounted eligible two-way interactions involve $m_2$ (skipping the already counted $m_1 m_2$ interaction), and so on up to $m-i$ new interactions for $m_i$. That is, there will be a total of $\sum_{j=1}^{i}(m-i) = mi - i(i+1)/2$ interactions eligible to be active for each set of $i$ active main effects. The remainder of the proof proceeds as above.

# Acknowledgments

I would like to thank Jeff Wu and Michael Hamada, who supervised this work as a Ph.D. thesis at the University of Waterloo. I am also grateful to Robert McCulloch, Merlise Clyde, and Giovanni Parmigiani for interesting discussions, and an anonymous referee for comments which improved the paper. This research was supported by the Natural Sciences and Engineering Research Council of Canada and the Manufacturing Research Corporation of Ontario.